# Increasing Data Equity Through Accessibility

Position statement, submitted to OSTP, October 2022

Signatories: Frank Elavsky, Carnegie Mellon University; Jennifer Mankoff, Co-Director Center for Research and Education on Accessible Technology and Experiences, University of Washington; Arvind Satyanarayan, MIT Visualization Group.

**Overview:** This response considers data equity specifically for people with disabilities[1]. The RFI asks "how Federal agencies can better support collaboration with other levels of government, civil society, and the research community around the production and use of equitable data." We argue that one critically underserved community in the context of data equity is people with disabilities. Today's tools make it extremely difficult for disabled people to (1) interact with data and data visualizations and (2) take jobs that involve working with and visualizing data. Yet access to such data is increasingly critical, and integral, to engaging with government and civil society. We must change the standards and expectations around data practices to include disabled people, and support the research necessary to achieve those goals.

We define *disability* here in terms of the discriminatory and often systemic problems with available infrastructure's ability to meet the needs of all people [UN 2017, Oliver, 2013]. We define *accessibility* as the creation of resources that can be accessed by anyone, regardless of their disability, with standard accessibility tools. We define *resources* as any data related analysis, production, and visualization platforms, as well as any data visualizations.

Data equity can level the playing field for people with disabilities both in opening new employment opportunities and through access to information, while data inequity may amplify disability by disenfranchising people with disabilities. Further, as Whittaker and colleagues (2019) state, "...discrimination against people of color, women, and other historically marginalized groups has often been justified by representing these groups as disabled…. Thus disability is entwined with, and serves to justify, practices of marginalization." (p. 11). Below we address three of the questions in the RFI that are most pertinent to the needs of disabled people: Questions 4, 5 and 6.

---

[1] We use the terms "people with disabilities" and "disabled people" and variations of them interchangeably in this document to reflect the varied choices that disabled people and theorists themselves choose.

## 4. What resources, programs, training, or other tools can expand opportunities for historically underrepresented scholars and research institutions to access and use equitable data across levels of government? 5. What resources, programs, training, or tools can increase opportunities for community-based organizations to use equitable data to hold government accountable to the American public?

We combine two questions into one here because we believe that the same underlying barriers impact access for scholars, research institutions, and community groups. We are focused specifically on scholars, research institutions, and community-based organizations that are underrepresented due to disability. We note that disability is not exclusive and can include anyone and at any phase of life. Thus, in addition to being a need in its own right, if we do not address disability access, we are further marginalizing underrepresented scholars and excluding research institutions and community-based organizations that serve any underrepresented population, due to the fact that their constituents will surely include people who are disabled *and* have other marginalized identities.

Our primary goal in answering this question is to highlight the opportunity to expand upon the government's use of accessible tools to produce accessible visualizations. While the government maintains a website containing accessibility information, which mentions data visualization,[2] there is an opportunity to expand this information and ensure that government workers are trained to make use of best practices[3] and the best tools[4] available when producing data visualizations. For example, while the United States Web Design System (USWDS) provides [some guidance on data visualization](), it is not sufficient for passing Section 508 or addressing the more complex access barriers involved in data visualization. Chartability, as an example, is an attempt to consolidate web standards, research, and practitioner knowledge towards a set of accessibility guidelines specific to the practice of visualizing and representing data and data interfaces (Elavsky, 2022).

Further, these standards must become integral to how the government produces data, across all areas of government. From the CDC to the Census Bureau, critical data that is highly important to all historically underrepresented peoples and should be available to underrepresented scholars and research institutions to access and use, must be accessible to fully include everyone.

## 6. What resources, programs, training, or tools can make equitable data more accessible and useable for members of the public?

In responding to this question, we will focus on two domains. First, as the question asks, we address *what can be done to make equitable data available to members of the public* with disabilities. However we note that Just having access to data is not enough, or just, when power, understanding and action are in the hands of government agents, computer scientists, business people and the many other stakeholders implementing data systems who do not themselves have disabilities. As Bennett and Keyes (2020) argue, we must look beyond fairness, which can only "reproduce the discrimination it seeks to remedy," to disability justice, a term used by activist scholars steeped in the Black Lives Matter movement (Wong, 2020). This means the government should also be able to equitably employ people with disabilities in jobs that produce equitable data. As such, the second half of our response to this question addresses *who can create equitable data resources.* Finally, we discuss the critical nature of *forward looking research investments* in advancing our ability to address both of these issues.

---

[2] https://designsystem.digital.gov/components/data-visualizations/
[3] https://chartability.fizz.studio/
[4] https://github.com/dataviza11y/resources

## 1. Improving the prevalence of accessible data for members of the public

We must expand the roles, education, laws, and processes around accessible data for current builders, makers, and practitioners. This is *the greatest area of impact for improving the accessibility of equitable data for members of the public.*

**Processes**: "Nothing about us without us" is a basic tenet of the disability rights movement (Charlton, 1998), and this applies in every domain. Thus, people with disabilities must be consulted when authoring policies that involve data, access, and equitable technology. Calls for information, involvement, and action should explicitly invite and encourage participation of those most affected.

**Roles**: Federal effort should be devoted to incentivizing affirmative action towards hiring and including people with disabilities in the US data workforce. This is critical to ensuring that data access is driven by and for people with disabilities, and guided by a deep understanding of how to meet the needs of this population. They must be a voice in the room when making decisions about data access.

**Education**: In some sense, we are all knowledge workers in today's world. Whether we are trying to understand election-related polling data or make informed decisions about COVID safety, we are interacting with data, often from a very young age. Thus, federal effort should be devoted to including technical accessibility topics and equitable data topics in K-12 curriculums. Similarly, higher education institutions, particularly those receiving government funding, should be expected to include accessibility topics in relevant curricula. Throughout the educational system, we should be training members of the public to have the skills to make use of accessible visualizations, and the skills and knowledge to produce them.

**Laws**: Given the critical nature of data for everything from making personal safety decisions to civic engagement, data access should be viewed as a human right. Thus, Section 508 (US DOJ, 1998) should be expanded from its current focus on federal contexts to enshrine access to information as an essential human right that applies to all domains where data and information are provided to the public and consumers.

**Tools:** The best visualizations we produce are often only as good as the tools we produce them with. We must find ways to lower barriers to producing accessible and equitable data. The tools that industry and local and state governments use are largely inadequate for producing robust, accessible data experiences. Our tools need to be improved. The [USWDS's Data Visualization](#) work should be expanded to include components that can be leveraged and not just guidelines. We note that a basic tenet of accessibility is that the *same resources* available to the general public should be accessible to people with disabilities -- it is not enough to simply place a data set on the web in an accessible tabular form, for example, and call that "equivalent" or "done". Time has shown that these parallel efforts frequently fall out of date, and rarely have the same features and capabilities of their original sources.

## 2. Including people with disabilities in the production of accessible data

Improving access to tools and improving the outcomes of our tools is one of the lowest-hanging areas of impact for addressing the scale of accessibility barriers in data

**Access to the means of production**: Currently, the production of accessible data visualizations is not readily available to the general public. This means that members of the public may sometimes request accessible visualizations (as when a blind student is provided with tactile graphics) or may sometimes encounter them (should a news report or government website follow best practices for accessible visualization production) but rarely have the means to create their own visualizations. An exception to this is the New York Public Libraries' [Dimension Lab](#), which provides access to accessible materials for the general public to make sense of data. Federal programs should be developed to ensure that such programs are widely available in public spaces across the nation.

**Improving tool accessibility**: On a related note, many data visualization platforms assume that people with disabilities (particularly people who are blind or low vision) will never themselves engage in the production of data visualizations. However, it is not only possible for people with disabilities to produce visualizations, it is productive and important due to the nature of visualization as a tool that itself serves other goals and learning (Potluri et al., 2022). As with any other technology, data visualization production tools should be expected, themselves, to adhere to best practices for accessibility.

## 3. Investing in an accessible, equitable data future

Anticipating future goals and challenges is an area sorely under-addressed in accessibility policy and strategy but will be vital to the success of equitable data in the future. Most efforts for accessibility are largely devoted to auditing and remediating based on minimum standards and guidelines, which consumes time, resources, and energy that should also be considered for longer-term goals.

**Research investment**: Funding must be provided for forward-thinking research that investigates structural and strategic limitations to equitable data access. More research is needed to investigate the ways that various cultural and socio-economic factors intersect with disability and access to technology; and to ensure that as data visualizations expand to include dashboards and other complex communication tools, our understanding of how to improve their accessibility and the tools to support that advance alongside. Further research is also needed to improve the accessibility of tools that are used to produce data visualizations.

**Community/local investment**: Funding must also be provided for community action and programs already in place, such as in schools, libraries, schools for the blind, in addition to programs that can be replicated from one context to another.

**Sharing**: In addition, the guidelines, tools, and findings between federal and non-federal efforts must be shared. Significant re-work takes place as initiatives are organized at the local level, such as the work from the [City of San Francisco's efforts](#) towards accessible public data services for COVID-19, when other localities embark on similar projects. This work could be systematically shared or organized at the federal level to improve how other cities, states, localities, or communities do this work as well as foster future collaborations.

## About the signatories

**Frank Elavsky**
Currently a PhD student at Carnegie Mellon University, researcher at Apple, and an invited expert and contributor to the Accessible Rich Internet Applications (ARIA) working group (a global accessibility guidelines working group, part of the W3C). Former lead engineer for Visa's accessible, open-source design system chart component library, [Visa Chart Components](#) and creator of [Chartability](#), a set of guidelines, examples, and heuristics for evaluating the accessibility of data visualizations.

**Jennifer Mankoff**
Prof. Mankoff is co-director of the [Center for Research and Education on Accessible Technology and Experiences](#). She is the Richard E. Ladner Professor in the Paul G. Allen School of Computer Science & Engineering at the University of Washington. Her research is focused on accessibility through giving people the voice, tools and agency to advocate for themselves.  She strives to bring both structural and personal perspectives to her work. She identifies as a person with a disability and her recent work has focused on accessible data access, including access to streaming data, data visualization dashboards, and new and more affordable means for the production of accessible tactile graphics. Her previous faculty positions include UC Berkeley's EECS department and Carnegie Mellon's HCI Institute. Jennifer is a CHI Academy member and has been recognized with an Alfred P. Sloan Fellowship, IBM Faculty Fellowship and SIGCHI Social Impact Award.

**Arvind Satyanarayan**
Prof. Satyanarayan is an Associate Professor of Computer Science in the Electrical Engineering and Computer Science (EECS) department at MIT, and a member of the MIT Computer Science and Artificial Intelligence Lab (MIT CSAIL). He leads the MIT Visualization Group, which uses data visualization as a petri dish to study intelligence augmentation (IA), or how computation can help amplify our cognition and creativity while respecting our agency. His group's recent work has focused on making data visualizations accessible to people with disabilities through screen reader technologies and natural language descriptions. Arvind's work has been recognized with an NSF CAREER award, a National Academy of Science Kavli Fellowship, an IEEE VGTC Significant New Researcher Award, and a Google Research Scholar Award. Visualization toolkits he has developed with collaborators are widely used in industry (including at Apple, Google, and Microsoft), on Wikipedia, and in the Jupyter data science community.